\documentstyle[12pt,fleqn]{article}
\newcommand{\be}{\begin{equation}}
\newcommand{\ee}{\end{equation}}
\newcommand{\bea}{\begin{eqnarray}}
\newcommand{\eea}{\end{eqnarray}}

\begin{document}
\title{Convergence properties of the cluster expansion
for equal-time Green functions in scalar theories
\footnote{supported by DFG, Forschungszentrum J\"ulich and GSI Darmstadt}}
\author{A. Peter, W. Cassing, J.M. H\"auser, and M.H. Thoma\\
Institut f\"ur Theoretische Physik, Universit\"at Gie\ss en\\
35392 Gie\ss en, Germany \\}
\maketitle
\date{\today}
\begin{abstract}
We investigate the convergence properties of the
cluster expansion of equal-time Green functions in scalar theories
with quartic self-coupling in $(0+1)$, $(1+1)$, and $(2+1)$ space-time
dimensions. The computations
are carried out within the equal-time correlation dynamics approach,
which consists in a closed set of coupled equations of motion
for connected Green functions as obtained by a truncation
of the BBGKY hierarchy.
We find that the cluster expansion shows good convergence
as long as the system is in a localized state
(single phase configuration) and that it
breaks down in a non-localized state (two phase configuration),
as one would naively expect.
Furthermore, in the case of dynamical calculations with a time
dependent Hamiltonian for the evaluation of
the effective potential we find two timescales determining the
adiabaticity of the propagation; these are the time required for
adiabaticity in the single phase region and the time required for
tunneling into the non-localized lowest energy state in the two
phase region. Our calculations show a good convergence for the
effective potentials in $(1+1)$ and $(2+1)$ space-time dimensions
since tunneling is suppressed in higher space-time dimensions.
\end{abstract}

\bigskip
\begin{center}
\end{center}
\newpage
\section{Introduction}
\label{introduction}
The many-body problem of quantum field theory, in spite of great
efforts in the past, still remains unsolved and there is a need
for genuine nonperturbative methods. In \cite{wan95} we have proposed
a connected (equal-time) Green function approach for SU(N) gauge
theories which might provide valuable insight into the low energy
QCD problem. However, the possible truncation schemes in the order
of the connected Green functions will be limited for practical
purposes and one thus needs intrinsic criteria that allow for a
judgement of the convergence properties of the approach especially
in those cases, where exact solutions are not available.

Before adressing the SU(N) Yang-Mills problem, we analyze the
convergence properties of the correlation dynamical approach for
the scalar quantum field theory with $\lambda \Phi^4$
self-interaction, which has become an important theoretical
laboratory for testing non-perturbative methods.
This is partly due to the fact that $\Phi^4$-theory
provides an easy framework for studying different scenarios in
nonperturbative renormalization when going from $(0+1)$
to $(1+1)$ \cite{cha75}-\cite{cea90}, $(2+1)$ \cite{ste85}-\cite{fre82}
and $(3+1)$ \cite{ste84}-\cite{hua87} space-time dimensions.

In two preceding articles \cite{hae95, pet96} we have
investigated ground state symmetry breaking and the
effective potential in $\Phi^4$-theory in
$(1+1)$ and $(2+1)$ space-time dimensions
using the equal-time correlation dynamics
method (in $(d+1)$-dimensional $\Phi^4$-theory denoted as
$\Phi^4_{d+1}CD$),
which consists in a coupled set of equations of motion
for connected equal-time Green functions. This coupled set of equations
is obtained by inserting the cluster expansion, i.e. the expansion
of full Green functions in terms of connected Green functions, into the
equations of the BBGKY hierarchy and by neglecting all connected
$n$-Point functions with $n>N$ for some given $N$.

The derivation of the correlation dynamical equations for $\Phi^4$-theory
has been presented in detail in \cite{hae95, pet96}; the
resulting equations in $(1+1)$ dimensions are given by Eqs. (20)-(31)
in appendix A of \cite{hae95} and in $(2+1)$ dimensions by
Eqs. (19)-(21) in Sect. 2 as well as Eqs. (28)-(36) in the appendix
of \cite{pet96}. The equations in $(0+1)$ dimensions can easily
be obtained from those in higher dimensions by discarding the
dependence on the spatial coordinates and
by omitting all mass counterterms.
For a more general discussion of
correlation dynamics in case of nonrelativistic fermionic systems
we refer the reader to refs. \cite{wan85, sch90, ghe93, pfi94}.

The calculations in \cite{hae95, pet96} aimed at the
evaluation of the effective potential $V_{eff}(\Phi_0)$,
which is given by the
minimum of the energy expectation value in the subspace of states
with a fixed expectation value of the scalar field
$\langle \Phi \rangle = \Phi_0$. Since the $\Phi^4_{d+1}CD$-equations
describe the (equal-time) evolution of the system for a
given initial condition, a direct (variational) evaluation of the
effectiv potential within this approach would amount to minimizing
the energy in the manifold of stationary solutions with the constraint
$\langle \Phi \rangle = \Phi_0$. In practice, however, this turns
out to be impossible because the energy is not
bounded from below without additional constraints, since the solution
manifold contains unphysical Green functions. Although e.g. for the
$(0+1)$-dimensional case in the 2-point approximation a sufficient
constraint can be derived from the uncertainty relation, it is not
yet clear what constraints for higher order approximations
have to be imposed.

The computation of the effective potential and the corresponding
lowest energy state Green functions within our approach
is therefore carried out by means of a dynamical calculation
using the Gell-Mann and Low theorem, i.e. by adiabatically
changing the parameters of the theory in time from a configuration
with a well known lowest energy state solution
to the actual configuration of interest
as described in \cite{hae95, pet96}.
Whereas this many-body scheme was found to converge quite rapidly
in case of nonrelativistic nuclear physics problems \cite{pfi94},
it is not clear if this also holds for the
quantum field theoretical case.
We thus have to explore the convergence
of the cluster expansion and its dependence on the parameters of
the theory, i.e. the coupling $\lambda$ and the expectation value of
the scalar field $\langle \Phi \rangle = \Phi_0$.

Before investigating $\Phi^4$-theory in $(1+1)$ and
$(2+1)$ space-time dimensions, we turn to the simpler case of
$(0+1)$ space-time dimensions, which is equivalent to the quantum
mechanical anharmonic oscillator in case of a positive squared mass
and to the quantum mechanical double-well potential in case of
a negative squared mass.
Since in $(0+1)$ dimensions we have
access to the exact lowest energy state solution for given
$\langle {\rm x} \rangle = {\rm x}_0$, we here can also directly
control the validity of the correlation dynamical solution.

This article is organized as follows: In Sect. \ref{nullpluseins} we
investigate the $(0+1)$-dimensional case in full detail with respect
to the problem of degenerate vacua and quantum tunneling, whereas
in Sect. \ref{einszweipluseins} we present our numerical
results for the $(1+1)$- and the $(2+1)$-dimensional case.
Sect. \ref{summary} concludes the present study with a brief
summary of our results, while technical details of the cluster
expansion as well as the GEP (Gaussian effective potential)
approximation are given in appendices \ref{clusterexpansion}
to \ref{gep}.
\section{Convergence of the cluster expansion in $(0+1)$ dimensions}
\label{nullpluseins}
In $(0+1)$ space-time dimensions and with the identification
$\Phi(t) \rightarrow {\rm x}(t)$, $\Pi(t) \rightarrow {\rm p}(t)$,
the Lagrangian of $\Phi^4$-theory with positive
squared mass $m^2$ (anharmonic oscillator) reads
\begin{equation}
L_1=\frac{1}{2} (\partial_t {\rm x})^2
-\frac{1}{2} m^2 {\rm x}^2
-\frac{1}{4} \lambda {\rm x}^4 \; ,
\label{nulleinsposlagrange}
\end{equation}
while the Hamiltonian is given by
\begin{equation}
H_1=\frac{1}{2} {\rm p}^2 + \frac{1}{2} m^2 {\rm x}^2
+ \frac{1}{4} \lambda {\rm x}^4
\label{nulleinsposhamilton}
\end{equation}
with the usual equal-time commutation relation
$[{\rm x}(t),{\rm p}(t)]=i$.

In case of a negative squared mass $m^2=-\mu^2$ (double-well potential)
we write the Lagrangian and the Hamiltonian as
\begin{equation}
L_2=\frac{1}{2} (\partial_t {\rm x})^2 + \frac{1}{2} \mu^2 {\rm x}^2
- \frac{1}{4} \lambda {\rm x}^4 - \frac{\mu^4}{4 \lambda}
=\frac{1}{2} (\partial_t {\rm x})^2 - \frac{1}{4} \lambda
({\rm x}^2-\frac{\mu^2}{\lambda})^2
\label{nulleinsneglagrange}
\end{equation}
and
\begin{equation}
H_2=\frac{1}{2} {\rm p}^2 +
\frac{1}{4} \lambda ({\rm x}^2 - \frac{\mu^2}{\lambda})^2
\; .
\label{nulleinsneghamilton}
\end{equation}

Since we are aiming at the effective potential,
i.e. the properties of the system not only as a
function of the coupling $\lambda$, but also as a function of
$\langle {\rm x} \rangle= {\rm x}_0$,
we explicitly split off the vacuum expectation value by writing
\begin{equation}
{\rm x}=x + {\rm x}_0 \; , \; \; \langle x \rangle = 0
\; ; \; \; {\rm p}=p
\label{splitoff}
\end{equation}
and treat ${\rm x}_0$ as a constant classical background field
which can be varied as an external parameter.
The explicit $\Phi^4_{0+1}CD$ equations then are obtained in the
usual way, i.e. by inserting the cluster expansion
(cf. appendix \ref{clusterexpansion}) into the coupled set of
equations of motion for the n-point Green functions
$\langle xx \rangle$, $\langle px \rangle$, $\langle pp \rangle$,
$\langle xxx \rangle$, $\langle pxx \rangle$, $\langle ppx \rangle$,
$\langle ppp \rangle$, etc. The resulting set of coupled equations
for the connected Green functions $\langle xx \rangle_c$,
$\langle px \rangle_c$, $\langle pp \rangle_c$, ..., then can be
closed by neglecting connected Green functions above some order
$n>N$, where $N$ is an integer $\ge 2$. Since the resulting set of
equations is quite lengthy and in case of $N=4$ has been presented
in refs. \cite{hae95, pet96} for $(1+1)$ and $(2+1)$ dimensions,
we omit an explicit representation here.

In order to investigate the convergence of the cluster expansion,
we compute the quantities
\begin{equation}
|\langle {\rm x}^n \rangle_c / \langle {\rm x}^n \rangle| \; ,
\label{nullpluseinsstaerke}
\end{equation}
where $\langle \cdot \rangle_c$ stands for the connected part of the
expectation value $\langle \cdot \rangle$ (cf. appendix \ref{xhochn})
for $n$ $=$ 2, 3, 4, 5, 6 in
the exact lowest energy state solution as well as in the $\Phi^4_{0+1}CD$
approach using a 6-point truncation scheme, i.e. including the connected
n-point functions with $n \le 6$. In line with the notations in
\cite{hae95, pet96} this set of equations is denoted
by $\Phi^4_{0+1}CD(2,3,4,5,6)$.

In our analysis we consider two quite distinct cases; i.e. an
extreme double-well potential with $\lambda/(4 \mu^3)=0.0158$ and
an anharmonic oscillator potential with $\lambda/(4 m^3)=10$.
The values (\ref{nullpluseinsstaerke})
of the exact solution have been
obtained by diagonalizing the Hamiltonian of the system with an
additional external source (cf. (\ref{nulleinsquellhamilton}))
in a sufficiently large set of basis states. The
$\Phi^4_{0+1}CD(2,3,4,5,6)$-results have been obtained
in the usual way by exploiting the Gell-Mann and Low theorem.
We start with the GEP solution \cite{ste85, hae95, tho92}
(cf. appendix \ref{gep}) and switch on the residual interaction
terms adiabatically. In case of the
anharmonic oscillator this corresponds to a time-dependent Hamiltonian
of the form
\begin{equation}
H(t)=\frac{1}{2} {\rm p}^2 + \frac{1}{2} M^2 {\rm x}^2
+ \frac{g(t)}{\lambda} \left[ \frac{m^2-M^2}{2} {\rm x}^2
+ \frac{1}{4} \lambda {\rm x}^4 \right] \; ,
\end{equation}
where the residual coupling
$g(t)$ runs from zero to $\lambda$, $M$ is the effective
mass of the GEP-solution and ${\rm x}_0$ is fixed. We recall that
in the correlation dynamical calculation the 1-point functions
$\langle x \rangle$ and $\langle p \rangle$ are set equal to zero
according to (\ref{splitoff}), and for each value
of ${\rm x}_0$ an individual time dependent calculation has to be
carried out.
We have used a linear time dependence of the form
$g(t)/4\mu^3=\beta t$ or $g(t)/4m^3=\beta t$, respectively, where
$\beta$ in each case has been chosen sufficiently small
such that the system shows adiabatic convergence
as long as ${\rm x}_0$ is
outside the domain where the exact solution indicates
a breakdown of the cluster expansion.
Our numerical results for (\ref{nullpluseinsstaerke}) in the lowest
energy state with given expectation value ${\rm x}_0$ are shown
in Fig. \ref{nulleinscorrelationstrength} as a function of
${\rm x}_0$ for both potentials. The explicit results from the
exact solution are displayed on the l.h.s. whereas the r.h.s. shows
the corresponding quantities in $\Phi^4_{0+1}CD(2,3,4,5,6)$
approximation. We note that
the sharp peaks for certain values of ${\rm x}_0/\sqrt{m}$ or
${\rm x}_0/\sqrt{\mu}$ in Fig. \ref{nulleinscorrelationstrength}
result from a change of sign of
$\langle {\rm x}^n \rangle_c / \langle {\rm x}^n \rangle$.

In the case of the double-well potential the exact solution
(upper left part of Fig. \ref{nulleinscorrelationstrength})
yields a dramatic sudden change in the relative importance of the
connected Green functions at a value of
${\rm x}^{crit}_0/\sqrt{\mu} \approx 3.8$.
Above ${\rm x}^{crit}_0/\sqrt{\mu}$ the relative importance of the connected
Green functions decreases by at least one order of magnitude as
one goes from $n$ to $n+1$, implying an excellent convergence of the
cluster expansion. In contrast to that, in the region
below ${\rm x}^{crit}_0/\sqrt{\mu}$ the full Green functions are dominated
by their connected parts, i.e. the system is dominated by fluctuations.

The corresponding result in $\Phi^4_{0+1}CD(2,3,4,5,6)$-approximation
(upper right part of Fig. \ref{nulleinscorrelationstrength})
above ${\rm x}^{crit}_0/\sqrt{\mu}$ shows nearly the same result as
the exact solution. This has been expected due to the excellent
convergence of the cluster expansion in this region. The sudden rise
in the relative importance of the higher order connected Green functions
in going from ${\rm x}_0/\sqrt{\mu} > {\rm x}^{crit}_0/\sqrt{\mu}$
to ${\rm x}_0/\sqrt{\mu} < {\rm x}^{crit}_0/\sqrt{\mu}$, however,
is smeared out over an extended region for the finite value of $\beta$
taken in the calculation. However, it becomes evident that
as soon as the connected Green functions dominate
the full Green functions (${\rm x}_0/\sqrt{\mu} \approx 2.84$), the method
of time dependently switching on the residual interaction within the
$\Phi^4_{0+1}CD(2,3,4,5,6)$-approximation breaks down.

In contrast to the results for the double-well potential, the ratios
(\ref{nullpluseinsstaerke})
for the anharmonic oscillator (lower two pictures in Fig.
\ref{nulleinscorrelationstrength}) demonstrate a good agreement between
the exact solution and the $\Phi^4_{0+1}CD(2,3,4,5,6)$-solution for all
${\rm x}_0/\sqrt{\mu}$.
The relative importance of the connected Green functions
increases with decreasing ${\rm x}_0/\sqrt{\mu}$, but even at
${\rm x}_0/\sqrt{\mu}=0$ the higher order full Green functions can in
a good approximation be expressed by their disconnected
parts (note that
$\langle {\rm x}^2 \rangle_c / \langle {\rm x}^2 \rangle \rightarrow 1$
as ${\rm x}_0 \rightarrow 0$ by definition).

In order to understand the numerical results for the double well
potential in
Fig. \ref{nulleinscorrelationstrength} in a more qualitative way,
we now examine the wave functions obtained by the exact solution
for the system.
We first note that instead of finding the lowest energy state in
the subspace of all states with given
$\langle {\rm x} \rangle = {\rm x}_0$, one can equivalently introduce
a source term into the Hamiltonian by writing
\begin{equation}
H^J_2=H_2-\mu J {\rm x}
\label{nulleinsquellhamilton}
\end{equation}
and then find the ground state of the system with respect to the
whole Hilbert space. The resulting state, which has some vacuum
expectation value ${\rm x}_0(J)$ depending on the source $J$, is
also the lowest energy state in the subspace with given ${\rm x}_0(J)$.

In Fig. \ref{wellenfunktion} we display the absolute square of the
ground state wavefunction $\Psi_0({\rm x})$ and the wavefunction of
the first excited state $\Psi_1({\rm x})$ of the Hamiltonian
(\ref{nulleinsquellhamilton}) with $\lambda/(4\mu^3)=0.0158$
as a function of $f(J)={\rm x}_0(J)/\sqrt{\mu}$
and ${\rm x}/\sqrt{\mu}$
where both wavefunctions have been obtained via diagonalization of
(\ref{nulleinsquellhamilton}). At $f(J)=0$, $|\Psi_0({\rm x})^2|$ and
$|\Psi_1({\rm x})^2|$ have
approximately the same shape with maxima at $\pm {\rm x}^{min}$.
We note that in the limit
$\lambda/(4\mu^3) \rightarrow 0$ the double-well potential
develops an infinitely high barrier at ${\rm x}=0$. This leads to a
spectrum of the Hamiltonian for $J=0$ which consists of pairs
of degenerate Eigenvalues.
Due to the infinite barrier at ${\rm x}=0$
there is no more tunneling between the
2 orthogonal states spanning the Eigenspace corresponding to
each Eigenvalue.
The lowest energy states of the system then are of the form
\begin{equation}
|\Psi_0\rangle=a |L_0\rangle + b |R_0\rangle
\label{mixphase}
\end{equation}
with $|L_0\rangle$ and $|R_0\rangle$ denoting the mutually orthogonal
lowest energy solutions
with $\langle L_0|{\rm x}| L_0 \rangle=-{\rm x}^{min}$
and $\langle R_0|{\rm x}| R_0 \rangle=+{\rm x}^{min}$,
i.e. the optimally localized ''left'' and ''right'' states which can be
constructed from an appropriate superposition of the degenerate
ground states. We then obtain
\begin{equation}
\langle \Psi_0 | {\rm x} |\Psi_0 \rangle = {\rm x}^{min}(|b|^2-|a|^2)
\; \in [ -{\rm x}^{min},{\rm x}^{min} ]
\end{equation}
with $\langle L_0| {\rm x} |R_0 \rangle=0$;
states of the form (\ref{mixphase}) thus restore the
convex shape of the effective potential by making it flat between
$-{\rm x}^{min}$ and ${\rm x}^{min}$.
As soon as $J \ne 0$, the system will energetically prefer
to shift its ground state wavefunction completely to
one side of the potential barrier, depending on the sign of $J$.
In other words, the system with an infinitely high potential barrier
undergoes a phase transition from the state $|L_0\rangle$ to the
state $|R_0\rangle$ as $J$ passes through zero.
Here we use the term ''phase transition'' for any
discontinuous behaviour of the lowest energy solution with respect
to the parameters of the Hamiltonian.

If $\lambda/(4\mu^3)$ assumes a finite (but still small) value
as in case of Fig. \ref{wellenfunktion},
the pairwise degeneracy of Eigenvalues disappears due to
tunneling -- which is no longer forbidden for finite $\lambda$ --
but the spectrum
still consists of pairs with a small energy gap.
The separation of the lowest Eigenvalues $\Delta E=E_1-E_0$
can be taken as a measure of the inverse tunneling time.
The region between $-{\rm x}^{min}$ and ${\rm x}^{min}$
then can be covered by a very small variation of $J$, i.e.
the above obtained ''phase transition'' from $|L_0\rangle$
to $|R_0\rangle$ at $J=0$ in case of
a finite coupling is smeared out into a sharp crossover.
As can be seen from Fig. \ref{wellenfunktion},
the ground state wavefunctions are (to a very good approximation) still
of the form (\ref{mixphase}), with only slight modifications due to
tunneling. The first excited state of the system then is
given by the Eigenstate corresponding to the higher
Eigenvalue in the lowest lying pair, i.e. approximately by
\begin{equation}
|\Psi_1\rangle=b^* |L_0\rangle - a^* |R_0\rangle \; .
\label{orthostate}
\end{equation}
In Fig. \ref{wellenfunktion}, this can be seen from the fact that
the relative strengths of the left
and right maxima of $|\Psi_1({\rm x})^2|$ and $|\Psi_0({\rm x})^2|$
depend on $f(J)$ in exactly the opposite way.

For $|f(J)| > {\rm x}^{min}/\sqrt{\mu}$ the system suddenly changes
its behaviour and the ground state and the first excited state are
localized states, which to a good approximation can be described as the
ground state and the first excited state of the harmonic oscillator
potential obtained in second order Taylor expansion of the
classical potential around its minima, as can also
be observed in Fig. \ref{wellenfunktion}.
This corresponds to the single phase localized
configuration at finite $J$ in the limiting case of the infinitely
high potential barrier.

We arrive at the conclusion that for a finite, but small
value of $\lambda/(4\mu^3)$ -- although
the system does not exhibit true ground state symmetry breaking
due to tunneling --,
it will still qualitatively behave like a system
which is in a (non-localized) two phase configuration for
$|{\rm x}_0| < {\rm x}^{min}$ and in a (localized) single phase
configuration for $|{\rm x}_0| > {\rm x}^{min}$.

It is now straightforward to show analytically, that the cluster
expansion has to break down in a two phase configuration according to
(\ref{mixphase}) as soon as it shows good convergence
in both optimally localized single phase configurations.
Furthermore, in a field theory
with an arbitrary space-time dimensionality,
the cluster decomposition property of the connected Green functions,
\begin{equation}
G^c_n(x_1,\cdots,x_p;y_{p+1},\cdots,y_n) \rightarrow 0
\quad \left( { min \atop {i=1,\cdots,p \atop j=p+1,\cdots,n}} |x_i-y_j|
\rightarrow \infty \right) \; ,
\end{equation}
is lost in a state constructed according to (\ref{mixphase})
whenever $|L_0\rangle$ and $|R_0\rangle$ are the optimally localized
states in case of a system with ground state symmetry breaking.
For the $(0+1)$ dimensional
system this has no direct implication for equal-time Green functions
since it only applies to the time arguments, e.g. in a time ordered
Green function.

Taking into account the above considerations, we can now identify
${\rm x}^{min}$, the expectation value of ${\rm x}$ in
the optimally localized right state,
with ${\rm x}^{crit}_0$, for which (\ref{nullpluseinsstaerke})
(cf. the upper left part of Fig. \ref{nulleinscorrelationstrength})
shows a relatively sharp rise.
In the correlation dynamical calculation (displayed in the
upper right part of Fig. \ref{nulleinscorrelationstrength}),
the sharp rise is smeared out. We also
observe that the region, in which the system is dominated by
fluctuations, only extends up to
${\rm x}_0/\sqrt{\mu} \approx 3.3$ instead of
${\rm x}_0/\sqrt{\mu} = {\rm x}^{crit}_0 \approx 3.8$
for the exact solution.

In order to get an explanation for this behaviour,
we show in Fig. \ref{idealmaxwell} a qualitative picture
of the external source $J$ as a
function of ${\rm x}_0$ for 3 different cases. In case (1)
(dashed line), there is no ground state
symmetry breaking and for each value of
$J$ there is a unique value ${\rm x}_0(J)$. In case (2), furthermore,
we have the situation corresponding to the convex shape of the
effective potential given by the construction of nonlocal states
(\ref{mixphase}) in case of
ground state symmetry breaking, where for any nonzero value of
$J$ there is a unique
value ${\rm x}_0(J)$ and for $J=0$ all values between
$-{\rm x}^{min}$ and ${\rm x}^{min}$ are accessible.
This corresponds to the field theoretical Maxwell construction
\cite{ami78}, where the two phases, given by the left and
the right optimally localized state in (\ref{mixphase}) ''coexist''.
In case (3)
we still have ground state symmetry breaking, but the curve for
$J({\rm x}_0)$ is analytically continued into the region, where
the field theoretical Maxwell construction takes place, implying
that no phase coexistence is allowed here.

Due to the non-ergodicity of a system, that posesses 2 phases
without tunneling in between,
the trajectory of case (3) is realized
in any time dependent adiabatic process as used in
our correlation dynamics approach for the computation of the effective
potential. This results in a good
convergence of the cluster expansion throughout the whole process,
because the system always stays in a localized state.
For the double-well potential with finite potential barrier
this is e.g. realized
within the $\Phi^4_{0+1}CD(2)$-limit, i.e. the
time-dependent Hartree-Fock method, because at this level
of approximation tunneling is not yet included.
The adiabatic limit of the $\Phi^4_{0+1}CD(2)$ solution
then simply is the GEP solution, i.e. the lowest energy Gaussian.

However, for a system with tunneling between the 2 phases we come to
the conclusion that the convergence properties of the cluster
expansion with respect to the adiabatic computation of the lowest
energy state
are determined by the relative magnitude of 2 timescales:
the specific tunneling time $t_{tun}$ and the
time $t_{prop}$ over which the system is propagated;
the latter quantity in our case is always chosen to ensure sufficient
adiabatic convergence in the single phase region.
The tunneling time for a given system itself depends on the correlation
dynamical truncation scheme; as indicated above, $t_{tun}$ is always
infinite within the 2-point truncation scheme
and the inclusion of higher order connected Green functions will
in general tend to decrease it.

If $t_{tun} \gg t_{prop}$, a dynamical phase mixing will be negligible
and our correlation dynamics will give useful results for the dynamical
time evolution of the system. In the two phase region, however,
the system will no longer propagate along an adiabatic trajectory of lowest
energy. Loosely speaking,
the correlation dynamical time evolution will
generate states of lowest energy in the ''subspace of sufficiently
localized states'', the latter of course not being mathematically
well defined.

On the other hand, for
$t_{tun} {\raisebox{-.5ex}{$\stackrel{<}{\scriptstyle\sim}$}} t_{prop}$
tunneling will lead to a breakdown of
correlation dynamics in the course of the time integration.
This is the reason for the breakdown of
the $\Phi^4_{0+1}CD(2,3,4,5,6)$ approach for
${\rm x}_0/\sqrt{\mu}
{\raisebox{-.5ex}{$\stackrel{<}{\scriptstyle\sim}$}} 3.3$
in the upper right part
of Fig. \ref{nulleinscorrelationstrength}.

In order to investigate the behaviour of our double-well system
with respect to the different time scales
in a manner directly related to the idealized picture given in
Fig. \ref{idealmaxwell}, we now choose a different dynamical
process, in which the background field ${\rm x}_0$ is changed
as a function of time. For the initialization at $t=0$ we
make use of the fact that for ${\rm x}_0 \rightarrow \infty$
(strong field limit) the system approaches the classical limit;
there again the GEP solution can be taken as a starting configuration
even though the residual interaction is completely taken into account
in the time propagation right from the beginning.

We then compute the corresponding
external source $J$ as a function of ${\rm x}_0$ by imposing the
Hellmann-Feynman theorem (which of course strictly only holds
for an exact Eigenstate of the system)
\begin{eqnarray}
\frac{d}{dJ} E_0(J)=\frac{d}{dJ} \langle \Psi_0(J) |
H^J_2 | \Psi_0(J) \rangle = \langle \Psi_0(J) | \frac{d}{dJ} H^J_2
|\Psi_0(J) \rangle \nonumber \\
= -\mu \langle \Psi_0(J)| {\rm x} | \Psi_0(J)
\rangle = -\mu {\rm x}_0(J) \; ,
\label{feynmanhellmann}
\end{eqnarray}
where $H^J_2$ is given by (\ref{nulleinsquellhamilton}).
Inserting (\ref{nulleinsquellhamilton}) into (\ref{feynmanhellmann})
yields
\begin{equation}
-\mu {\rm x}_0 (J) = \frac{d}{dJ} \langle H^J_2 \rangle =
\frac{d}{dJ} \langle H_2 \rangle - \mu {\rm x}_0(J)
-\mu J \frac{d}{dJ} {\rm x}_0(J) \; .
\end{equation}
With
\begin{equation}
\frac{d}{dJ} \langle H_2 \rangle = \left( \frac{d}{d{\rm x}_0}
\langle H_2 \rangle \right) \left( \frac{d}{dJ} {\rm x}_0(J) \right)
\end{equation}
we obtain
\begin{equation}
J({\rm x}_0)=\frac{1}{\mu} \frac{d}{d{\rm x}_0} \langle H_2 \rangle
\end{equation}
relating the external source $J$ to the effective potential in the
usual way. The philosophy behind this definition of $J({\rm x}_0)$
within a dynamical calculation is the interpretation already indicated
above, that for $t_{tun} \gg t_{prop}$ the system will
''quasi-adiabatically'' evolve in the subspace of localized
(single phase) states.

In Fig. \ref{realmaxwell} the external source $J$ is plotted
as a function of ${\rm x}_0/\sqrt{\mu}$ in
$\Phi^4_{0+1}CD(2,3,4,5,6)$-approximation for a double-well
potential with $\lambda/(4\mu^3)=0.0158$, where ${\rm x}_0$ has been
changed in time according to ${\rm x}_0(t)/\sqrt{\mu}
={\rm x}^{start}_0/\sqrt{\mu} -\alpha t$.
The fat solid line is the result in $\Phi^4_{0+1}CD(2)$-approximation
with $\alpha=0.00001 \; {\rm c/fm}$ ($\mu^2=1 \; {\rm MeV}^2$),
which essentially lies on top of the
GEP result and is shown for comparison.
For large values of $\alpha$ the trajectory in the $(J-{\rm x}_0)$--plane
can be computed down to ${\rm x}_0=0$ without any problem and exhibits
the same qualitative behaviour as the trajectory of case (3) in
Fig. \ref{idealmaxwell}, indicating that for
$\alpha > 0.01024 \; {\rm c/fm}$
we have $t_{tun} \gg t_{prop} \propto 1/\alpha$.

By reducing $\alpha$ we get into the region where
$t_{tun} \approx t_{prop}$. The inclusion of higher order connected
Green functions then allows for tunneling, thus leading to a breakdown
of the dynamical time propagation due to the dominance of
fluctuations in the phase coexistence regime. With decreasing $\alpha$,
the value of ${\rm x}_0$, at which this breakdown
happens, shifts towards the point where $J$ changes sign
(${\rm x}_0 \approx 3.8$).
In the exact solution this value corresponds to the boundary of the
phase coexistence region (cf. upper left part of Fig.
\ref{nulleinscorrelationstrength}).
\section{Convergence of the cluster expansion in $(1+1)$ and $(2+1)$
dimensions}
\label{einszweipluseins}
We now turn to the technically much more involved case of $(1+1)$-
and $(2+1)$-dimensional $\Phi^4$-theory. Both theories are
superrenormalizable, and only a mass renormalization is required.
The Lagrangian and the Hamiltonian are given by
\begin{equation}
{\cal L}=\frac{1}{2} \partial_\mu \Phi \partial^\mu \Phi
-\frac{1}{2} m_0^2 \Phi^2 - \frac{1}{4} \lambda \Phi^4 \; ,
\label{einszweilagrange}
\end{equation}
\begin{equation}
H=\frac{1}{2} \int d^\nu x \left[ \Pi^2 + ( \nabla \Phi )^2
+ m_0^2 \Phi^2 + \frac{1}{2} \lambda \Phi^4 \right] \; ,
\label{einszweihamilton}
\end{equation}
where $m_0^2=m^2 + \delta m^2$ is the bare mass, $m^2$ the renormalized
mass and $\nu$ denotes the number of spatial dimensions.
$\delta m^2$ contains the mass counterterms, which in $(2+1)$
dimensions are given by the tadpole diagram and the setting sun diagram,
and in $(1+1)$ dimensions by the tadpole diagram only (cf.
\cite{hae95, pet96}).

The field operators and their conjugate momenta are expanded
into plane waves in a box with periodic boundary conditions
according to
\begin{equation}
\Phi(x)=\sum_\alpha \Phi_\alpha \psi_\alpha(x) \; ,
\quad
\Pi(x)=\sum_\alpha \Pi_\alpha \psi_\alpha(x)
\label{basisexpansion}
\end{equation}
with $\psi_\alpha(x)=\frac{1}{\sqrt{V}}e^{i\vec{k}_\alpha \vec{x}}$;
$V=L^\nu$ and $\vec{k}_\alpha$ chosen according to the boundary
conditions.

For our numerical simulations in $(1+1)$ dimensions we use
$L=100 \;{\rm fm}$,
$m=10 \; {\rm MeV}$ and the 21 lowest lying
plane waves; in $(2+1)$ dimensions we use $L=20 \; {\rm fm}$,
$m=10 \; {\rm MeV}$ and the 29 lowest lying states.
For further details of the explicit set of equations solved we
refer the reader to refs. \cite{hae95, pet96}.

In Fig. \ref{einszweieinscorrelationstrength} we show the quantities
\begin{equation}
\left| \sum_{\alpha \beta}
\langle \Phi_{\alpha} \Phi_{\beta} \rangle_c
\left. \right/ \sum_{\alpha \beta}
\langle \Phi_{\alpha} \Phi_{\beta} \rangle
\right|
=\left| \langle \Phi^2(x) \rangle_c \left. \right/
\langle \Phi^2(x) \rangle \right|
\; (n=2) \;,
\label{stark2}
\end{equation}
\begin{equation}
\left| \sum_{\alpha \beta \gamma}
\langle \Phi_{\alpha} \Phi_{\beta} \Phi_{\gamma} \rangle_c
\left. \right/ \sum_{\alpha \beta \gamma}
\langle \Phi_{\alpha} \Phi_{\beta} \Phi_{\gamma} \rangle
\right|
=\left| \langle \Phi^3(x) \rangle_c \left. \right/
\langle \Phi^3(x) \rangle \right|
\; (n=3)
\label{stark3}
\end{equation}
and
\begin{equation}
\left| \sum_{\alpha \beta \gamma \delta}
\langle \Phi_{\alpha} \Phi_{\beta} \Phi_{\gamma} \Phi_{\delta} \rangle_c
\left. \right/ \sum_{\alpha \beta \gamma \delta}
\langle \Phi_{\alpha} \Phi_{\beta} \Phi_{\gamma} \Phi_{\delta} \rangle
\right|
=\left| \langle \Phi^4(x) \rangle_c \left. \right/
\langle \Phi^4(x) \rangle \right|
\; (n=4) \; ,
\label{stark4}
\end{equation}
as a function of $\Phi_0= \langle \Phi(x) \rangle
= L^{-\nu/2} \langle \Phi(\vec{k}=\vec{0}) \rangle$.
Note that due to translation invariance all Green functions with
$\sum_{\alpha_i} \vec{k}_{\alpha_i} \ne \vec{0}$ have to vanish.
The quantities (\ref{stark2}), (\ref{stark3}) and (\ref{stark4})
have been chosen
as a measure for the relative importance of the higher order connected
Green functions in analogy to (\ref{nullpluseinsstaerke}).
All results have been obtained within the
$\Phi^4_{\nu+1}CD(2,3,4)$-approximation
(i.e. including everything up to the connected 4-point function),
which is the lowest order in correlation dynamics to which the
setting sun diagram contributes, which diverges logarithmically in
(2+1) dimensions (cf. \cite{pet96}).
The coupling has been switched on in time according to
\begin{equation}
\frac{\lambda}{4 m^{3-\nu}}=\beta t
\end{equation}
with the free (perturbative vacuum) solution
as initialization.
This method is more efficient than switching
on the residual interaction with the GEP solution as an initialization,
since it yields results for the lowest energy state over a whole
range of coupling constants in only a single time dependent run.
In the $(0+1)$-dimensional case, the method of switching on the
residual interaction (leading to a correlated state for only one
value of the coupling constant) had to be chosen, because the
double well potential has no
''unperturbed phonon state'', because for zero coupling the
potential is a harmonic oscillator with negative mass and thus
not bounded from below.
The vacuum expectation value
$\Phi_0$ has been split off as a classical
background field and the 1-point function has not been
propagated explicitly in analogy to the treatment of
${\rm x}_0$ in the $(0+1)$-dimensional case.
The value $\beta=0.05 \; {\rm c/fm}$ has been chosen
small enough in order to observe an adiabatic behaviour of the
time dependent solution.

The results for $(1+1)$ dimensions are shown in the left column and
those for $(2+1)$ dimensions in the right column of
Fig. \ref{einszweieinscorrelationstrength} for various values
of the coupling constant.
As in Fig. \ref{nulleinscorrelationstrength}, the peaks result
from taking the absolute value of a quantity that changes sign
and have no physical meaning.
In both, $(1+1)$ and $(2+1)$ dimensions,
the relative importance of the connected
3- and 4-point functions increases with increasing coupling and
decreasing $\Phi_0$; however, for all parameters considered this
relative importance decreases in going from $n$ to $n+1$, and moreover,
the highest contribution of the connected 3-point function to
the full 3-point function is only about 20 $\%$.
The cluster expansion
thus shows satisfactory convergence throughout the investigated
parameter range.

The critical couplings for a transition into the
symmetry broken phase in $(1+1)$ and $(2+1)$ dimensions
within the $\Phi^4_{\nu+1}CD(2,3,4)$-approximation are
(for the present specific choice of numerical parameters
specifying the plane wave basis set)
$\lambda/4m^2=2.247$ and $\lambda/4m=0.369$, respectively
\cite{hae95, pet96}. Below these couplings a good convergence
of the cluster expansion can be expected, since the system is
in a single phase, i.e. it can be described by a localized wave
functional. The fact, that we still observe this convergence
above the critical coupling indicates that the time $t_{tun}$
required for building up a two phase configuration via tunneling
within the $\Phi^4_{\nu+1}CD(2,3,4)$-approximation is infinite
or at least large
compared to the time $t_{prop}$ needed for the ''quasiadiabatic''
propagation of the system along the lowest energy localized state
trajectory in the sense of Sect. \ref{nullpluseins}. In the
latter case, the system would (over a certain range of
$t_{prop} \propto 1/\beta$) seem to converge against an asymptotic
configuration as $t_{prop}$ increases, and then suddenly change
its behaviour when tunneling kicks in.
We have only been able to observe such a behaviour
in the $(0+1)$ dimensional case, but within the investigated
range of $\beta$-values not in $(1+1)$ and $(2+1)$ dimensions.

In fact, it is well known that tunneling is suppressed as the number
of space-time dimensions increases. In $(1+1)$ dimensions, by means of
a soliton-antisoliton pair, it can be shown that tunneling between the
two vacua with $\Phi_0=\pm \Phi_0^{min}$ is still possible \cite{ste85}.
In $(2+1)$ dimensions, however, no tunneling is expected \cite{ste85};
this would be the lowest space-time dimension with
real ground state symmetry breaking, i.e. with an effective potential
which is flat between $-\Phi_0^{min}$ and $+\Phi_0^{min}$.

A simple qualitative argument \cite{ste85} pointing
towards the suppression of
tunneling in higher space-time dimensions is the following:
Consider a bubble of phase 1 within a region of phase 2. The
surface region of this bubble then has a higher energy density than
the surrounding. In $(1+1)$ dimensions, the surface region is
given by the environment of the two end-points of the bubble
interval (the soliton and the antisoliton)
and its size is therefore independent of the size of the
bubble; consequently the bubble can expand without increasing the
total energy of the system.
Generally, in $(d+1)$ dimensions, the size of the
surface region behaves as $R^{d-1}$, so that for $d \ge 2$
the bubble cannot expand without a corresponding increase in
the total energy density. The above geometrical considerations
are similar to those used for the proof of Derrick's theorem
\cite{der64}
(non-existence of topological solitons in scalar theories with
$d \ge 2$), where a rescaling of the classical field according to
$\Phi(x) \rightarrow \Phi(\Lambda x)$ is used to show that any
potential topological soliton solution would be energetically
unstable against variation of $\Lambda$.
\section{Summary}
\label{summary}
In this work we have explored the convergence
properties of the cluster expansion approach and presented criteria
within the theory itself, that allow to conclude about its
convergence also in those cases, where the exact solution is not
known. As an example we have studied scalar theories with quartic
self coupling in $(0+1)$, $(1+1)$, and $(2+1)$ dimensions.
We have emphasized the aspect of a nonperturbative computation
of the lowest energy state with given field expectation value,
since our previous work \cite{hae95,pet96} aimed
at the evaluation of the effective potential.

The $\Phi^4$ theory in $(0+1)$
dimensions is equivalent to a quantum mechanical anharmonic
oscillator for positive squared mass and a quantum mechanical
double-well potential for negative squared mass.
In the case of an anharmonic oscillator with $\lambda/(4m^3)=10$ we
obtained a good convergence of the cluster expansion as well as an
excellent agreement between the exact solution
and the correlation dynamical
solution; this could be explained by the fact that the system is in
a single phase (localized state) configuration.

In case of an extreme double-well potential with
$\lambda/(4\mu^3)=0.0158$ the cluster expansion
in the exact solution breaks down as soon as the ground state
for a given expectation value $\langle {\rm x} \rangle$
can approximately be described by a Maxwell construction, i.e.
by a coherent superposition of two single phase configurations.
The shrinking of the
region with dominant fluctuations within the correlation dynamical
calculation could be traced back to the fact, that the time $t_{prop}$,
chosen according to the requirement of an
adiabatic propagation in the single phase region,
is small compared to the tunneling time $t_{tun}$ of the system
for certain values of $\langle {\rm x} \rangle$.

Our analysis for the QFT systems in $(1+1)$ and $(2+1)$ dimensions
demonstrated, that even in the phase coexistence
(non-localized state) region, i.e.
for coupling constants higher than the
critical coupling for ground state symmetry breaking,
the correlation dynamical approach yields
a good convergence of the cluster expansion. Here,
the tunneling times within our dynamical calculation
are infinite or at least
very large compared to the time required for an adiabatic
propagation; this behaviour is traced back to the fact that
tunneling is suppressed in $(1+1)$ dimensions and
probably absent in $(2+1)$ dimensions.

The main result of this article can therefore be summarized as follows:
the applicability of correlation dynamics (which is based on an
expansion in terms of connected Green functions) to the adiabatic
computation of the lowest energy state with given field expectation
value is given if a) the exact stationary solution is a single phase,
and b) if the required propagation time for adiabatic
or ''quasiadiabatic'' (cf. Sect. \ref{nullpluseins}) convergence
is small compared to the tunneling time.
This is in fact
the case in higher space-time dimensions since there tunneling
is progressively suppressed. This is not only important for
lowest energy state calculations as carried out in this article,
but also for applications of correlation dynamics to
the nonequilibrium dynamics of systems with phase transitions.
Furthermore, the relative ratios (\ref{stark2}-\ref{stark4})
provide an intrinsic test of the correlation dynamical method also
for those field theoretical problems, where the exact solution is
not known.
\begin{appendix}
\section{The cluster expansion}
\label{clusterexpansion}
The explicit form of the cluster expansion can be derived from
the generating functionals of full and connected Green functions
(cf. e.g. \cite{hae95}),
$Z[J,\sigma]$ and $W[J,\sigma]$, given by
\begin{equation}
Z[J,\sigma]={\rm Tr} \left\{ \rho T \left[ e^{i \int d^{\nu+1}x
(J(\hat{x})\Phi(\hat{x}) + \sigma (\hat{x}) \Pi(\hat{x}))} \right]
\right\}
\end{equation}
and
\begin{equation}
Z[J,\sigma]=e^{W[J,\sigma]} \; ,
\end{equation}
respectively, where $T$ is the time ordering operator, $\rho$
is the statistical density operator describing the state of the system
(${\rm Tr} \rho=1$) and $\hat{x}$ denotes $(x,t)$.
We start with the cluster expansions for the time-ordered Green
functions:
\begin{eqnarray}
{\langle}\Phi(\hat{x}) {\rangle} \;=\; {\langle}\Phi(\hat{x})
{\rangle}_c
\; \; , \; \; \; \;
{\langle}\Pi(\hat{x}) {\rangle} \;=\; {\langle}\Pi(\hat{x})
{\rangle}_c \; ,
\end{eqnarray}
\begin{eqnarray}
\lefteqn{
{\langle}T \Phi(\hat{x}_1) \Phi(\hat{x}_2){\rangle} =
\lim_{J, \sigma \to 0}
\frac{\delta}{i\delta J(\hat{x}_1)}
\frac{\delta}{i\delta J(\hat{x}_2)} e^{W[J, \sigma]}
} \nonumber \\
& & = \lim_{J, \sigma \to 0}
\frac{\delta}{i\delta J(\hat{x}_1)} \left\{ \left(
\frac{\delta}{i\delta J(\hat{x}_2)} W[J, \sigma] \right)
e^{W[J, \sigma]}
\right\}
\nonumber \\
& & = \lim_{J, \sigma \to 0} \left\{ \left(
\frac{\delta}{i\delta J(\hat{x}_1)}
\frac{\delta}{i\delta J(\hat{x}_2)} W[J, \sigma] \right)
+ \left( \frac{\delta}{i\delta J(\hat{x}_1)} W[J, \sigma] \right)
\left( \frac{\delta}{i\delta J(\hat{x}_2)} W[J, \sigma] \right)
\right\}
e^{W[J, \sigma]}
\nonumber \\
\nonumber \\
& & = {\langle}T \Phi(\hat{x}_1) \Phi(\hat{x}_2){\rangle}_c +
{\langle}\Phi(\hat{x}_1){\rangle}{\langle}\Phi(\hat{x}_2)
{\rangle} \; ,
\end{eqnarray}
where $\langle \cdot \rangle_c$ denotes the connected part of the
expectation value.
Analogously we obtain
\begin{eqnarray*}
\langle T \Pi(\hat{x}_1) \Phi(\hat{x}_2) \rangle = \langle
T \Pi(\hat{x}_1)
\Phi(\hat{x}_2) \rangle_c
+ \langle \Pi(\hat{x}_1) \rangle \langle \Phi(\hat{x}_2) \rangle \; ,
\end{eqnarray*}
\begin{eqnarray*}
\langle T \Pi(\hat{x}_1) \Pi(\hat{x}_2) \rangle = \langle
T \Pi(\hat{x}_1)
\Pi(\hat{x}_2) \rangle_c
+ \langle \Pi(\hat{x}_1) \rangle \langle \Pi(\hat{x}_2) \rangle \; ,
\end{eqnarray*}
\begin{eqnarray*}
\lefteqn{
{\langle}T \Phi(\hat{x}_1) \Phi(\hat{x}_2) \Phi(\hat{x}_3)
{\rangle} \; = \;
{\langle}T \Phi(\hat{x}_1) \Phi(\hat{x}_2) \Phi(\hat{x}_3)
{\rangle}_c
+ {\langle}T \Phi(\hat{x}_1) \Phi(\hat{x}_2){\rangle}_c
{\langle}\Phi(\hat{x}_3)
{\rangle}
} \nonumber \\
& & + {\langle}T \Phi(\hat{x}_1) \Phi(\hat{x}_3){\rangle}_c
{\langle}\Phi(\hat{x}_2){\rangle}
+ {\langle}T \Phi(\hat{x}_2) \Phi(\hat{x}_3) {\rangle}_c
{\langle}\Phi(\hat{x}_1){\rangle}
+ {\langle}\Phi(\hat{x}_1){\rangle} {\langle}\Phi(\hat{x}_2){\rangle}
{\langle}\Phi(\hat{x}_3){\rangle}
\; , \; \; ... \; \; .
\end{eqnarray*}
\begin{eqnarray}
\end{eqnarray}
The expressions for equal-time Green functions are obtained by
taking the
well-defined equal-time limit which yields the appropriate
operator ordering
in the cluster expansions. We arrive at
\begin{eqnarray*}
{\langle}\Phi(x) {\rangle} \;=\; {\langle}\Phi(x){\rangle}_c
\; \; , \; \; \; \;
{\langle}\Pi(x) {\rangle} \;=\; {\langle}\Pi(x){\rangle}_c \; ,
\end{eqnarray*}
\begin{eqnarray*}
{\langle} \Phi(x_1) \Phi(x_2){\rangle}
= {\langle} \Phi(x_1) \Phi(x_2){\rangle}_c + {\langle}\Phi(x_1)
{\rangle}{\langle}\Phi(x_2){\rangle} \; ,
\end{eqnarray*}
\begin{eqnarray*}
\langle  \Pi(x_1) \Phi(x_2) \rangle = \langle  \Pi(x_1) \Phi(x_2)
\rangle_c
+ \langle \Pi(x_1) \rangle \langle \Phi(x_2) \rangle \; ,
\end{eqnarray*}
\begin{eqnarray*}
\langle  \Pi(x_1) \Pi(x_2) \rangle = \langle  \Pi(x_1) \Pi(x_2)
\rangle_c
+ \langle \Pi(x_1) \rangle \langle \Pi(x_2) \rangle \; ,
\end{eqnarray*}
\begin{eqnarray*}
\lefteqn{
{\langle} \Phi(x_1) \Phi(x_2) \Phi(x_3) {\rangle} \; = \;
{\langle} \Phi(x_1) \Phi(x_2) \Phi(x_3) {\rangle}_c
+ {\langle} \Phi(x_1) \Phi(x_2){\rangle}_c {\langle}\Phi(x_3){\rangle}
} \nonumber \\
& & + {\langle} \Phi(x_1) \Phi(x_3){\rangle}_c {\langle}\Phi(x_2)
{\rangle}
+ {\langle} \Phi(x_2) \Phi(x_3) {\rangle}_c {\langle}\Phi(x_1)
{\rangle}
+ {\langle}\Phi(x_1){\rangle} {\langle}\Phi(x_2){\rangle}
{\langle}\Phi(x_3){\rangle}
\; , \; \; ... \; \; ,
\end{eqnarray*}
\begin{eqnarray}
\end{eqnarray}
where all (equal) time arguments have been suppressed.
In view of their length the cluster expansions for the other Green
functions required for our calculations are not explicitly given here,
but can be found in \cite{wan95}.
The explicit forms of the cluster expansions of
$\langle {\rm x}^n \rangle$, $n=1,\cdots,6$ (required for our
analysis in $(0+1)$ dimensions) are given in appendix \ref{xhochn}.
\section{Cluster expansion of $\langle {\rm x}^n \rangle$}
\label{xhochn}
Applying the derivation of appendix \ref{clusterexpansion},
we obtain the following expressions for the cluster
expansion of $\langle {\rm x}^n \rangle$, $n=1,\cdots,6$
in $\Phi^4_{0+1}$-theory using $\Phi(x) \rightarrow {\rm x}$:
\begin{eqnarray}
\langle {\rm x} \rangle = \langle {\rm x} \rangle_c
\; ,
\end{eqnarray}
\begin{eqnarray}
\langle {\rm x}^2 \rangle = \langle {\rm x} \rangle_c^2
+ \langle {\rm x}^2 \rangle_c
\; ,
\end{eqnarray}
\begin{eqnarray}
\langle {\rm x}^3 \rangle = \langle {\rm x} \rangle_c^3
+ 3 \langle {\rm x} \rangle_c \langle {\rm x}^2 \rangle_c
+ \langle {\rm x}^3 \rangle_c
\; ,
\end{eqnarray}
\begin{eqnarray}
\langle {\rm x}^4 \rangle = \langle {\rm x} \rangle_c^4
+ 6 \langle {\rm x} \rangle_c^2 \langle {\rm x}^2 \rangle_c
+ 3 \langle {\rm x}^2 \rangle_c^2
+ 4 \langle {\rm x} \rangle_c \langle {\rm x}^3 \rangle_c
+ \langle {\rm x}^4 \rangle_c
\; ,
\end{eqnarray}
\begin{eqnarray}
\langle {\rm x}^5 \rangle = \langle {\rm x} \rangle_c^5
+ 10 \langle {\rm x} \rangle_c^3 \langle {\rm x}^2 \rangle_c
+ 15 \langle {\rm x} \rangle_c \langle {\rm x}^2 \rangle_c^2
+ 10 \langle {\rm x} \rangle_c^2 \langle {\rm x}^3 \rangle_c
+ 10 \langle {\rm x}^2 \rangle_c \langle {\rm x}^3 \rangle_c
\nonumber \\
+ 5 \langle {\rm x} \rangle_c \langle {\rm x}^4 \rangle_c
+ \langle {\rm x}^5 \rangle_c
\; ,
\end{eqnarray}
\begin{eqnarray}
\langle {\rm x}^6 \rangle = \langle {\rm x} \rangle_c^6
+ 15 \langle {\rm x} \rangle_c^4 \langle {\rm x}^2 \rangle_c
+ 45 \langle {\rm x} \rangle_c^2 \langle {\rm x}^2 \rangle_c^2
+ 15 \langle {\rm x}^2 \rangle_c^3
+ 20 \langle {\rm x} \rangle_c^3 \langle {\rm x}^3 \rangle_c
\nonumber \\
+ 60 \langle {\rm x} \rangle_c \langle {\rm x}^2 \rangle_c
\langle {\rm x}^3 \rangle_c
+ 10 \langle {\rm x}^3 \rangle_c^2
+ 15 \langle {\rm x} \rangle_c^2 \langle {\rm x}^4 \rangle_c
+ 15 \langle {\rm x}^2 \rangle_c \langle {\rm x}^4 \rangle_c
+ 6 \langle {\rm x} \rangle_c \langle {\rm x}^5 \rangle_c
+ \langle {\rm x}^6 \rangle_c
\; .
\end{eqnarray}
\section{The GEP approximation in $(0+1)$ dimensions}
\label{gep}
We consider the Hamiltonian
\begin{equation}
H=\frac{1}{2} {\rm p}^2 + \frac{1}{2} m^2 {\rm x}^2
+ \frac{1}{4} \lambda {\rm x}^4
= \frac{1}{2} p^2 + \frac{1}{2} m^2 (x+{\rm x}_0)^2
+ \frac{1}{4} \lambda (x+{\rm x}_0)^4
\; ,
\label{nulleinsposhamilton}
\end{equation}
where $m^2$ can be positive or negative.
The GEP (Gaussian effective potential) in $(0+1)$ dimensions is
given by a constrained variational calculation with a
Gaussian trial wave function, i.e. by the ansatz
\begin{equation}
\Psi_{{\rm x}_0}(x)=\left( \frac{M}{\pi} \right)^{\frac{1}{4}}
e^{-\frac{M}{2} x^2} \; .
\label{quasivac}
\end{equation}
This ansatz can then be seen as the ground state of a
variational harmonic Hamiltonian of the form
\begin{equation}
H=\frac{1}{2} p^2 + \frac{1}{2} M^2 x^2
\end{equation}
with an effective mass $M$. The connection
to the field theoretical Hartree-Fock-Bogoliubov ansatz becomes
obvious by noting, that (\ref{quasivac}) is the quasiparticle vacuum
with respect to Bogoliubov-transformed creation and
annihilation operators.
Keeping this last point in mind, it is also straightforward to see
why in a Gaussian trial state all connected Green functions of
higher order than the two-point function vanish; this property
follows directly from Wick's theorem.

The resulting equation for the effective mass is
\begin{equation}
M^2-m^2=3 \lambda \left( {\rm x}_0^2+\frac{1}{2M} \right)
\; ,
\end{equation}
which is equivalent to the Gap equation
following from the Bogoliubov-ansatz.
\end{appendix}
\newpage
\newpage
{\large \bf Figure Captions}
\newcounter{figno}
\begin{list}%
{\underline{fig.\arabic{figno}}:}%
{\usecounter{figno}\setlength{\rightmargin}{\leftmargin}}
\item
\label{nulleinscorrelationstrength}
The ratios
$|\langle {\rm x}^n \rangle_c / \langle {\rm x}^n \rangle|$ for
$n$=2,...,6 as a function of
${\rm x}_0/\sqrt{\mu}=\langle {\rm x} \rangle/\sqrt{\mu}$
for the double-well potential
with $\lambda/(4 \mu^3)=0.0158$ (upper two pictures) and for the
anharmonic oscillator potential with $\lambda/(4 m^3)=10$
(lower two pictures). L.h.s.: exact results; r.h.s.: results
within the $\Phi^4_{0+1}CD(2,3,4,5,6)$-approximation obtained by
time dependently switching on the residual interaction.
\item
\label{wellenfunktion}
Absolute squares of the ground state wave function $\Psi_0({\rm x})$
and the wavefunction of the first excited state $\Psi_1({\rm x})$
(both obtained from the exact solution)
as a function of ${\rm x}$ and of
$f(J)=\langle \Psi_0| {\rm x} |\Psi_0 \rangle$ in the case of a
double-well potential with $\lambda/(4\mu^3)=0.0158$ and an external
source $J$.
\item
\label{idealmaxwell}
Qualitative picture of the external source $J$ as a function of
${\rm x}_0=\langle {\rm x} \rangle$ for 3 different cases:
(1) no symmetry breaking, (2) Maxwell construction in case of
symmetry breaking, (3) analytical continuation into the two phase
region of the Maxwell contruction in case of symmetry breaking.
\item
\label{realmaxwell}
External source $J$ as a function of ${\rm x}_0/\sqrt{\mu}$ in
$\Phi^4_{0+1}CD(2,3,4,5,6)$-approximation for a double-well
potential with $\lambda/(4\mu^3)=0.0158$; ${\rm x}_0$ is changed
in time according to ${\rm x}_0(t)/\sqrt{\mu}
={\rm x}^{start}_0/\sqrt{\mu} -\alpha t$.
The fat solid line is the result in $\Phi^4_{0+1}CD(2)$-approximation
with $\alpha=0.00001 \; {\rm c/fm}$ ($\mu^2=1 \; {\rm MeV}^2$).
\item
\label{einszweieinscorrelationstrength}
The ratios
$\left| \sum_{\alpha \beta}
\langle \Phi_{\alpha} \Phi_{\beta} \rangle_c
\left. \right/ \sum_{\alpha \beta}
\langle \Phi_{\alpha} \Phi_{\beta} \rangle
\right|$ $(n=2)$,
$\left| \sum_{\alpha \beta \gamma}
\langle \Phi_{\alpha} \Phi_{\beta} \Phi_{\gamma} \rangle_c
\left. \right/ \sum_{\alpha \beta \gamma}
\langle \Phi_{\alpha} \Phi_{\beta} \Phi_{\gamma} \rangle
\right|$ $(n=3)$ and
$\left| \sum_{\alpha \beta \gamma \delta}
\langle \Phi_{\alpha} \Phi_{\beta} \Phi_{\gamma} \Phi_{\delta} \rangle_c
\left. \right/ \sum_{\alpha \beta \gamma \delta}
\langle \Phi_{\alpha} \Phi_{\beta} \Phi_{\gamma} \Phi_{\delta} \rangle
\right|$ $(n=4)$
as a function of $\Phi_0=\langle \Phi(x) \rangle$
for $(1+1)$ (l.h.s.) and $(2+1)$ (r.h.s.) dimensions in
$\Phi^4CD(2,3,4)$-approximation; all results have been obtained by
switching on the coupling adiabatically.
\end{list}
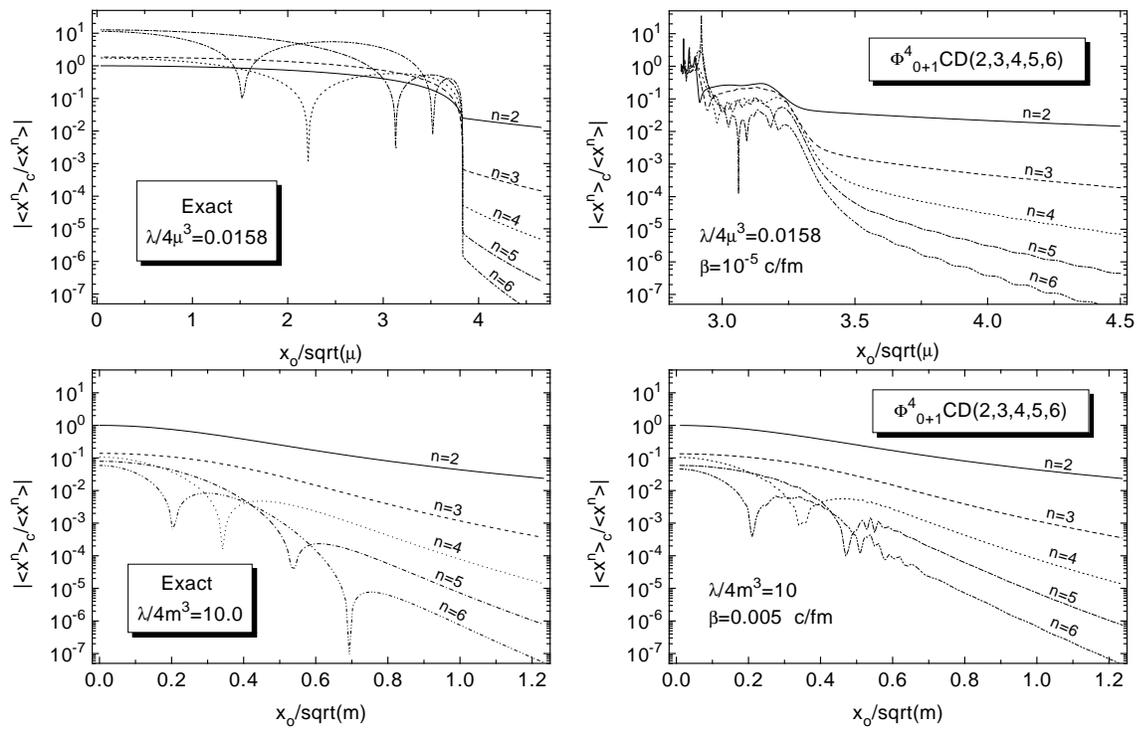
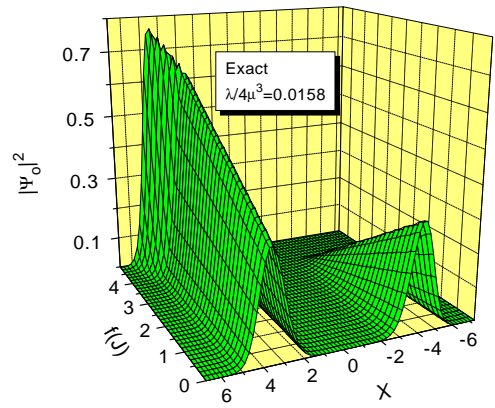
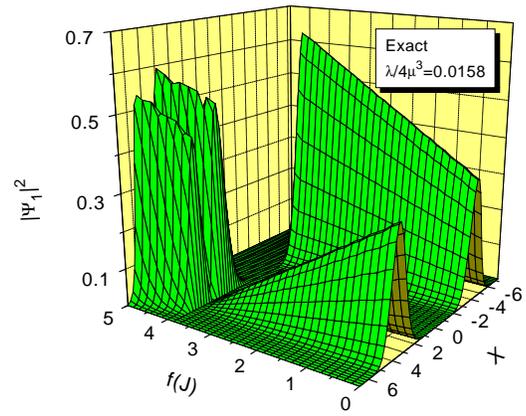
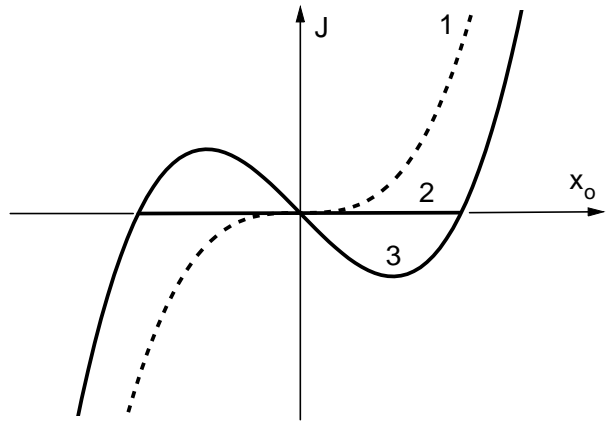
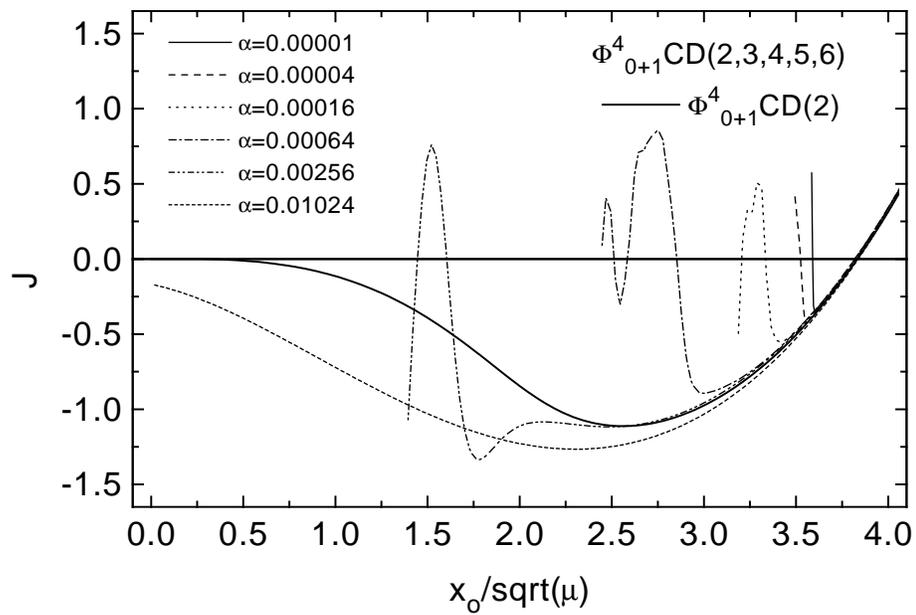
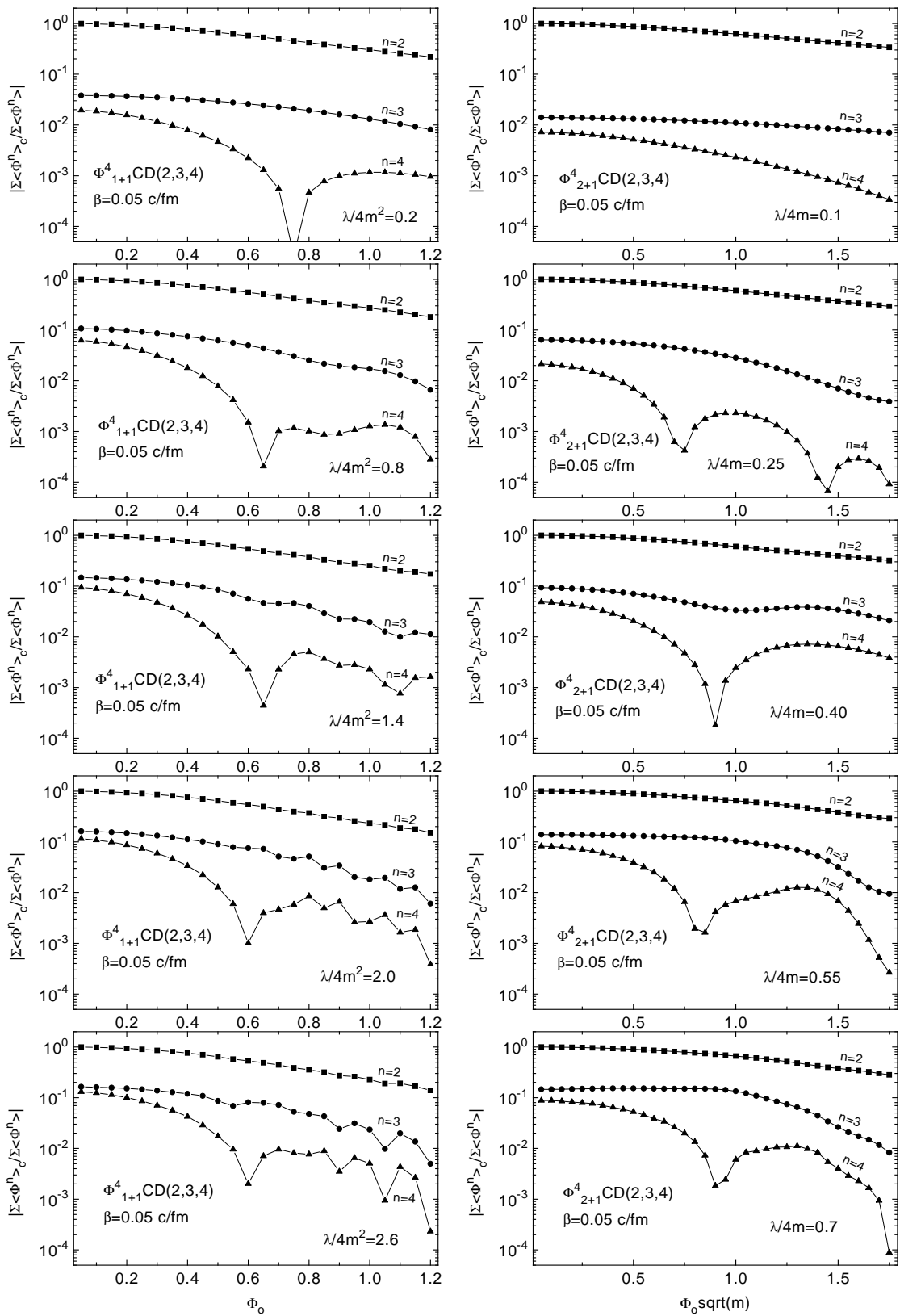
\end{document}